# Robust spin-polarized midgap states at step edges of topological crystalline insulators

Paolo Sessi,[1*] Domenico Di Sante,[2] Andrzej Szczerbakow,[3] Florian Glott,[1] Stefan Wilfert,[1] Henrik Schmidt,[1] Thomas Bathon,[1] Piotr Dziawa,[3] Martin Greiter,[2] Titus Neupert,[4] Giorgio Sangiovanni,[2] Tomasz Story,[3] Ronny Thomale,[2] and Matthias Bode[1,5]

[1]Physikalisches Institut, Experimentelle Physik II, Universität Würzburg, Am Hubland, 97074 Würzburg, Germany.

[2]Institut für Theoretische Physik, Universität Würzburg, Am Hubland, 97074 Würzburg, Germany.

[3]Institute of Physics, Polish Academy of Sciences, Aleja Lotników 32/46, 02-668 Warsaw, Poland.

[4]Physik-Institut, Universität Zürich, Winterthurerstrasse 190, CH-8057 Zürich, Switzerland.

[5]Wilhelm Conrad Röntgen-Center for Complex Material Systems (RCCM), Universität Würzburg, Am Hubland, 97074 Würzburg, Germany.

*Correspondence to: sessi@physik.uni-wuerzburg.de.

**Abstract**: Topological crystalline insulators are materials in which the crystalline symmetry leads to topologically protected surface states with a chiral spin texture, rendering them potential candidates for spintronics applications. Using scanning tunneling spectroscopy, we uncover the existence of one-dimensional (1D) midgap states at odd-atomic surface step edges of the three-dimensional topological crystalline insulator (Pb,Sn)Se. A minimal toy model and realistic tight-binding calculations identify them as spin-polarized flat bands connecting two Dirac points. This non-trivial origin provides the 1D midgap states with inherent stability and protects them from backscattering. We experimentally show that this stability results in a striking robustness to defects, strong magnetic fields, and elevated temperature.

**Main Text:**

The recent theoretical prediction and experimental realization of topological insulators (TIs) have considerably extended the notion of a phase of matter. Within this framework, it has been shown that—based on some topological invariants—the electronic properties of materials can be classified into distinct topological classes (*1,2*). In topologically non-trivial materials, unconventional boundary modes have been experimentally detected by several different techniques (*3–9*). In two-dimensional (2D) TIs, counter-propagating spin-momentum–locked one-dimensional (1D) edge modes develop along the sample boundary; in contrast, three-dimensional (3D) TIs (*4*) have boundary modes that are linearly dispersing chiral surface states. Although a large variety of 3D TIs have been reported, only very few 2D TIs are known [HgTe (*3*), InAs (*10*) quantum wells, and Bi bilayers (*11*)]. These 2D TIs are delicate and difficult to realize experimentally because they all require the fabrication of precisely controlled thin film heterostructures. Properties such as small band gaps (3,10), strong substrate-induced hybridization effects (11), or the existence of residual trivial states (10,11) make helical edge states not only challenging to study, but also of limited appeal for applications. Furthermore, their topological properties are protected only as long as time-reversal symmetry is preserved.

Here we report that two-dimensional (2D) topological surfaces, in turn, can be the mother state for non-trivial one-dimensional (1D) midgap states, suggesting a dimensional hierarchy of boundary states in topological insulators. Specifically, we report on the discovery of 1D topological spin-filtered channels that naturally develop at step edges of 3D topological crystalline insulators (TCIs), i.e., materials where the existence of surface Dirac states is guaranteed by crystal symmetries.

Figure 1A displays the rock-salt structure of $Pb_{1-x}Sn_xSe$ ($x \leq 0.4$). Depending on Sn-content $x$, these compounds have been reported to belong to two topologically distinct phases (*13*) that can be stoichiometrically controlled: starting from PbSe, which is topologically trivial, the substitutional solid solution $Pb_{1-x}Sn_xSe$ turns into a topologically non-trivial phase as the Sn concentration exceeds $x \approx 0.24$ (*14,15*). Its bulk inverted band gap cannot be adiabatically connected to a trivial state as long as some crystal symmetries are preserved. The electronic properties of high-symmetry surfaces of these TCIs are characterized by topologically protected linearly dispersing Dirac states (*13,14,16-19*). Figure 1B illustrates this scenario for the non-polar (001) surface, which is commonly exposed after cleaving bulk crystals. It hosts four Dirac cones centered in close proximity to the $\bar{X}$ and $\bar{Y}$ points of the Brillouin zone. Figure 1C shows a typical image of the (001) surface acquired by scanning tunneling microscopy (STM) on a freshly cleaved $Pb_{0.67}Sn_{0.33}Se$ bulk crystal (*20*), i.e. a material safely within the topological regime at $x \geq 0.24$. An atomically resolved image showing the Se sublattice (*21*) is displayed as an inset. The profile taken along the grey line shows that several atomically flat terraces exist, separated by step edges of different heights.

Given the equal probability of breaking the crystal bonds at every atomic layer, all steps can be mapped onto two different classes: those corresponding to an integer number $n$ of the conventional unit cell heights $a$, i.e. $na$, from here on even steps, and those corresponding to half integer multiples, i.e. $(1/2 + n)a$, called odd steps hereafter Whereas even steps maintain the translational symmetry of the surface lattice, odd steps introduce a structural $\pi$ shift (Figs. 1, D and E, show the two situations from the top). As will be described below, this has far reaching implications for the electronic structure of the step edges.

Figure 1F shows a differential conductance d$I$/d$U$ map at an energy of $E - E_\text{F} = -75$ meV, which was measured simultaneously with the topographic image presented in Fig. 1C. Whereas the local density of states (DOS) is similar for all terraces and also remains essentially unchanged for the even step edge, a strong enhancement can be recognized along the odd step. Its intensity is symmetrically distributed on both sides of the step and has a width of approximately 10 nm (line profile in Fig. 1F). Remarkably, this DOS is very homogeneously distributed along the step; no evidence for scattering-induced quasiparticle interference pattern was found (for the Fourier-transformed d$I$/d$U$ intensity along the step edge see Ref. 20, Fig. S1).

In order to unravel the origin of this prominent electronic feature, we measured local scanning tunneling spectroscopy (STS) data on the three atomically flat terraces and the two step edges. The precise locations, where the STS curves displayed in Fig. 1G have been obtained, are indicated by numbers 1–5 in Fig. 1F. As expected, all spectra obtained on top of Pb$_{0.67}$Sn$_{0.33}$Se(001) terraces (panels 1, 3, and 5) show the same features. In particular, similar to earlier observations on TCI surfaces (*17,22*), the minimum visible at approximately −75 meV marks the position of the Dirac point; the peaks L$^-$ and L$^+$ identify van Hove singularities, which signal the energy position of the two saddle points below and above the Dirac point, respectively.

Whereas this local DOS profile remains essentially unperturbed at the even step edge [position (2)], completely different STS curves are observed at odd step edges, e.g., at position (4). Here the spectrum is characterized by a strong peak at the Dirac point. More generally, its emergence is associated with a strong redistribution of the spectral weight over a relatively large energy range, as evidenced by the disappearance of the features associated to the van Hove singularities. This scenario is consistently also found for even and odd steps of higher order, i.e., triple and quadruple step edges (see Ref. 20, Fig. S2). Furthermore, the absence of

scattering[similar to recent experiments on weak topological insulators (23)], the energy position locked at the Dirac point, and the association with translational symmetry breaking all point to a topological origin of these 1D channels.

To unequivocally prove that the emergence of these 1D edge states is linked to the existence of a non-trivial bulk band structure, we have performed measurements on crystals where the Sn concentration has been systematically changed, thereby spanning the entire range from trivial to topological surfaces (14) (Fig. 2). For all three concentrations, i.e. (A,B) $x = 0$, (C,D) $x = 0.24$, and (E,F) $x = 0.33$, the topographic images displayed in Fig. 2, A, C, and E show one even and one odd step edge (see respective line profiles at the bottom of each panel). Irrespective of the step's even- or oddness, no particular edge feature is visible in the d$I$/d$U$ map of the topologically trivial material, i.e. of pure PbSe without any Sn (Fig. 2B). An edge state with a slightly enhanced d$I$/d$U$ intensity develops once $x = 0.24$ (Fig. 2D), i.e. just at the critical Sn concentration required for the transition between a normal to an inverted band gap, thereby leading to a weakly protected topological non-trivial state (14). By further increasing the amount of Sn to $x = 0.33$, the Pb$_{1-x}$Sn$_x$Se compound is driven deeply into the topological regime. Correspondingly, we observe a strong enhancement of the local d$I$/d$U$ signal at the position of the odd step edge (Fig. 2F).

To understand the accumulation of midgap states at odd step edges, it is best to think of them as electronic domain walls in the surfaces states (24). This effective 2D electronic surface state features four Dirac cones, and domain walls are created by interchanging the two atoms in the lattice unit cell. The simplest 2D toy model displaying an even number of Dirac cones and a two-sublattice structure is the staggered flux model, i.e., a square lattice model restricted to nearest-neighbor hopping terms, in which neighboring plaquets are threaded by opposite magnetic fluxes $+\phi$ and $-\phi$, with $0 < \phi < \pi$. We model a domain wall by interchanging the relative position of the two sublattices along the [11] direction (Fig. 3A). On the domain wall, the 2D Brillouin zone (BZ) of the staggered flux lattice is projected onto a 1D BZ, as indicated in green in Fig. 3B. Thereby, the two Dirac cones are not projected onto each other. We solve this system by Fourier transformation in the $q_\parallel$ direction only, while keeping a real space index for the direction perpendicular to the wall. We obtain a chain model with alternating, real hopping amplitudes $t_\pm = 2\cos\left(\frac{q_\parallel}{\sqrt{2}} \pm \frac{\phi}{4}\right)$, with $-\frac{\pi}{\sqrt{2}} < q_\parallel \leq \frac{\pi}{\sqrt{2}}$, and a domain wall (Fig. 3C). Except for the positions of the two Dirac cones in the 1D BZ, $q_\parallel = 0$ and $q_\parallel = \frac{\pi}{\sqrt{2}}$, where $|t_+| = |t_-|$, the 1D model is gapped. Because closing the chain in the presence of the domain wall requires an odd number of sites, whereas particle–hole symmetry demands that all states with $E \neq 0$ are arranged in pairs with energies $\pm E$, we necessarily obtain a midgap state at $E = 0$ localized around the domain wall. We obtain one such midgap state for each value of $q_\parallel \neq 0, \frac{\pi}{\sqrt{2}}$, i.e., one for each real space unit cell. The spectrum will hence schematically look as depicted in Fig. 3D.

To elevate this minimal model result to an accurate description of the experimental scenario observed in Fig. 1, we have performed tight-binding calculations adapted to reproduce the band-structure of a topologically non-trivial crystalline insulator (*13*). The model is solved for a geometry with two step edges on the topmost surface; the bottom surface is left unperturbed. Figure 3E shows the calculated band structure of an infinite terrace (left panel), a single-, a double-, and a triple-step edge (right) as a function of $q_\parallel$, the momentum along the step edge. In close analogy to the staggered flux model described above, we consider the case of step edges parallel to the [110] direction. A discussion of the step edges along the [100] direction (as those shown in the sketch of Fig. 1D and E), as well as further details of the model and the calculation can be found in (*20*). In all cases, the electronic structure obtained for the step-edge-free surface corresponds to the results expected for a TCI surface, i.e., with two Dirac cones symmetrically shifted with respect to $q_\parallel = \pi/a$ (grey lines). The size of the red dots is proportional to the degree of localization at the step edge.

Whereas even step edges give rise to two Dirac cones and a relatively wide open mouth-shaped feature connecting them, the odd-edge spectrum is significantly different. The states in between the Dirac cones have a narrow dispersion and therefore account for the overwhelming part of the step edge spectral weight (20). In contrast to the result of the staggered flux model presented above, these narrow states are not completely flat because the tight-binding model is not perfectly particle-hole symmetric. Furthermore, the midgap states do not extend over the whole Brillouin zone, as the Dirac cones are shifted away from high-symmetry points. For odd step edges, these states are laterally (i.e. perpendicular to the step) confined to a few nm, whereas in the z-direction they extend up to four lattice spacings. As shown in the left inset of Fig. 3E, our results for the single-atomic step edge correctly reproduce the spin-momentum locking of electronic states in the Dirac cone with an in-plane spin polarization. The right inset reveals that the two pairs of narrow, counter-propagating in-gap bands exhibit instead a high degree of out-of-plane spin polarization (maximum expectation value ≈ 0.7 for the operator $\langle \sigma_z \rangle$).

In comparison to 2DTIs the $Pb_{1-x}Sn_xSe$ topological edge state reported here is expected to offer superior properties as it is protected by a bulk band gap which is up to 200 meV wide (*14*). We experimentally tested the robustness of the edge state to perturbations, such as hybridization with adjacent edge states, external magnetic field, or enhanced temperatures. For example, Fig. 4A shows a topographic STM image of a sample location with four odd step edges. Whereas the two right step edges run almost in parallel thereby maintaining their separation, the two left step edges are inclined with respect to one another and eventually merge close to the bottom of the image. As qualitatively evidenced by the d$I$/d$U$ map of Fig. 4B and quantitatively supported by the line sections plotted at the bottom of this panel, the edge state disappears as soon as the step–step separation decreases below the spatial extent of the edge state (*25*), i.e., about 10 nm. We further analyze the response of these edge states to high magnetic fields *B*. Figure 4C reports a d$I$/d$U$ map (top) and STS data acquired on an odd step edge (bottom) at $B = 11$ T; contrary to the quantum spin Hall state found in HgTe, the 1D TCI state investigated here is robust against time-reversal symmetry breaking perturbations. Finally, Fig. 4D shows that the edge state also persists at elevated temperatures ($T = 80$ K). Despite the reduced intensity evidenced by the STS spectrum, a well-defined 1D channel is still clearly present.

The observation of a distinct type of one-dimensional states at odd step edges of topological crystalline insulators with relatively wide bulk band gaps opens up opportunities for the utilization of topological materials for sensing and information processing purposes well beyond existing materials (*4,10,11*). Furthermore, the absence of scattering and the high degree of spin polarization observed in tight-binding calculations indicate that the one-dimensional midgap state might be useful for spintronics applications. By patterning the step-and-terrace structure of TCI surfaces this may allow for the creation of well-separated conductive channels with a width of about 10 nm only. This may lead to interconnections between functional units at ultra-high packing densities. To fully explore whether the one-dimensional midgap state found at odd TCI step edges display quantum conductance effects, further investigations by e.g. four-probe transport measurements will be needed that address single step edges.

**Acknowledgments:** This research was supported by DFG (through SFB 1170 "ToCoTronics"; projects A02, B04, and C05) and by the Polish National Science Centre NCN grants 2014/15/B/ST3/03833 and 2012/07/B/ST3/03607. We further acknowledge support by the European Research Council (ERC) through ERC-StG-TOPOLECTRICS-Thomale-336012. DDS and GS gratefully acknowledge the Gauss Centre for Supercomputing e.V. (www.gauss-centre.eu) for funding this project by providing computing time on the GCS Supercomputer SuperMUC at Leibniz Supercomputing Centre (LRZ, www.lrz.de).

**Supplementary Materials:**

Materials and Methods

SupplementaryText

Figs. S1 to S7

References (26, 27)

FIG. 1. **Electronic properties of $Pb_{0.67}Sn_{0.33}Se$ terraces and step edges probed by STS.** (**A**) Rock-salt crystal structure and (**B**) schematic band structure of (Pb,Sn)Se. (**C**) Topographic STM image of a cleaved $Pb_{0.67}Sn_{0.33}Se$ surface (scan parameters: $U = -75$ mV; $I = 50$ pA; T=?). Inset: atomic resolution image of the Se sublattice. Two steps are visible in the main panel. The line section (bottom panel) measured along the grey line shows that their heights correspond to a single- (right) and a double-atomic step (left), respectively. Whereas the periodicity is maintained for even step edges (**D**), odd step edges lead to a structural $\pi$ shift (**E**). (**F**) d$I$/d$U$ map (top) measured at the same location as C. The line section (bottom) reveals an enhanced conductance at the position of the single-atomic step edge. (**G**) Local tunneling spectra measured with the STM tip positioned at the locations indicated in F. The spectra measured on atomically flat terraces (1, 3, and 5) and at even step edges (2) display the typical V-shape with a minimum at the Dirac energy ($E_D = -75$ meV) surrounded by two maxima indicating van Hove singularities ($L^- = -110$ meV and $L^+ = -30$ meV); the spectrum measured at the position of the odd step (4) exhibits a strong peak at the Dirac energy.

FIG. 2. **Sn concentration-dependent electronic properties of (Pb,Sn)Se.** Topography (left), d$I$/d$U$ maps (right), and their corresponding profiles taken along the indicated line (bottom of each panel), measured on $Pb_{1-x}Sn_xSe$ crystals with different Sn content, i.e. (**A**),(**B**) $x = 0$, (**C**),(**D**) $x = 0.24$, and (**E**),(**F**) $x = 0.33$, thereby spanning the range from trivial to topological surfaces. Step edges on the trivial compound ($x = 0$) carry no particular edge feature, irrespective of their even- or oddness. In contrast, a weak and strong enhancement of the local DOS is indicated by the high d$I$/d$U$ signal measured at odd step edges for $x = 0.24$ and $x = 0.33$, respectively. Scan parameters: $U = -310$ mV, $I = 30$ pA ($x = 0$); $U = -115$ mV, $I = 50$ pA ($x = 0.24$); $U = -70$ mV, $I = 100$ pA ($x = 0.33$). T= 4.8 K.

FIG. 3. **Theoretical analysis of the TCI step edge electronic structure**. (A) Staggered flux lattice with a domain wall in the [11] direction. (B) Brillouin zone including the projection onto the 1D domain wall. (C) Effective 1D hopping model describing the staggered flux lattice. (D) Schematic view of the resulting spectrum. (E) Tight-binding results of the electronic structure (momentum $q_\parallel$ parallel to the edge) for a 3D TCI with [110]-oriented step edges of different height. The continuum of bulk states and surface contributions are shown as grey lines and red dots, respectively. Whereas strongly localized narrow bands connecting the two Dirac cones emerge at odd step edges, no such spectral weight is observed for even step edges. The two insets show the spin polarization in the presence of a single-atomic step edge: in-plane component perpendicular to the step edge ($x$; left) and the out-of-plane component ($z$; right) with the color (red and blue) indicating opposite spin directions.

FIG. 4. **Robustness of the edge state against perturbations**. (A) Topographic STM image showing a $Pb_{0.67}Sn_{0.33}Se$ surface area exhibiting a total of four odd step edges. The separation between the two leftmost step edges continuously reduces from the top to the bottom of the image. (B) The $dI/dU$ map reveals that the intensity of the edge state fades away as the edges approach below about 10 nm. Line sections along the lines indicated by numbers are plotted in the bottom panels. $dI/dU$ maps taken around odd steps at (C) $B = 11$ T and (D) $T = 80$ K show the robustness of the 1D edge state against perturbations.

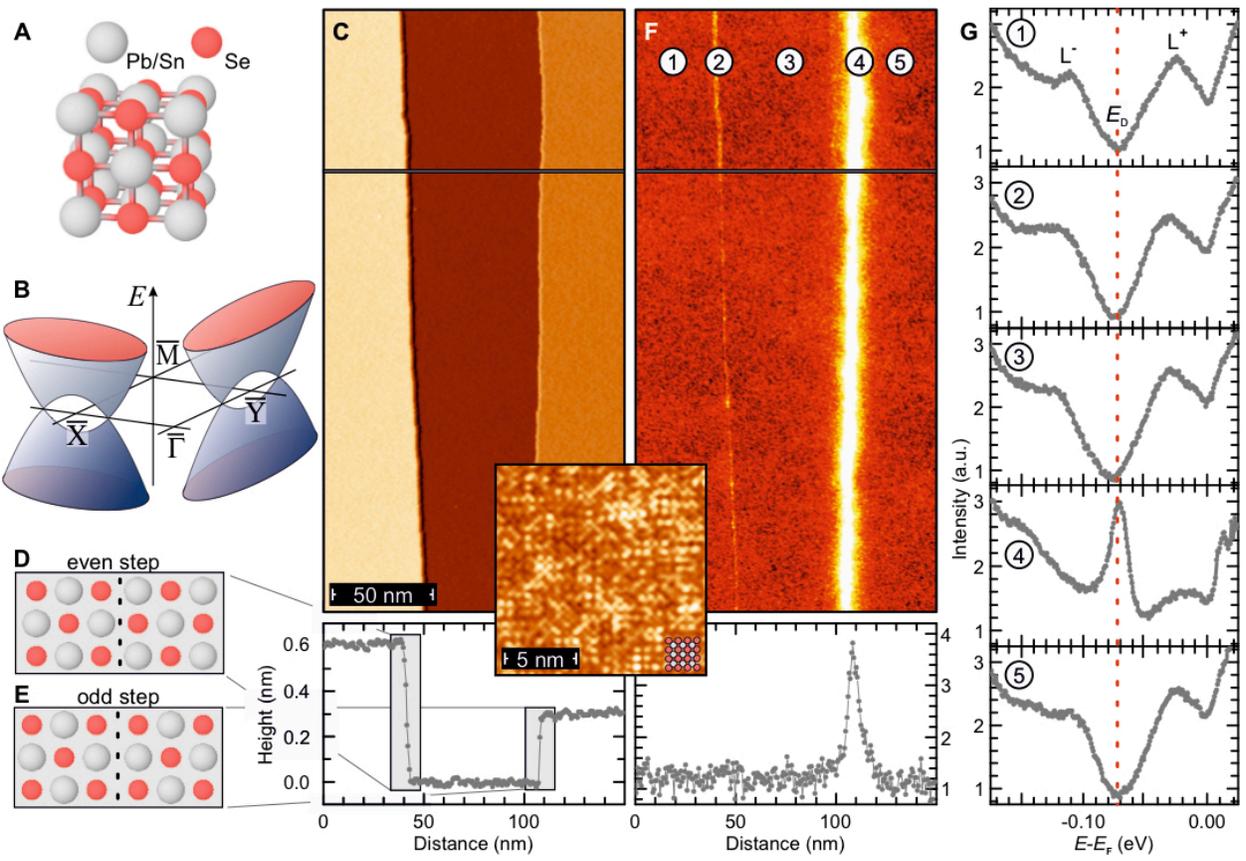

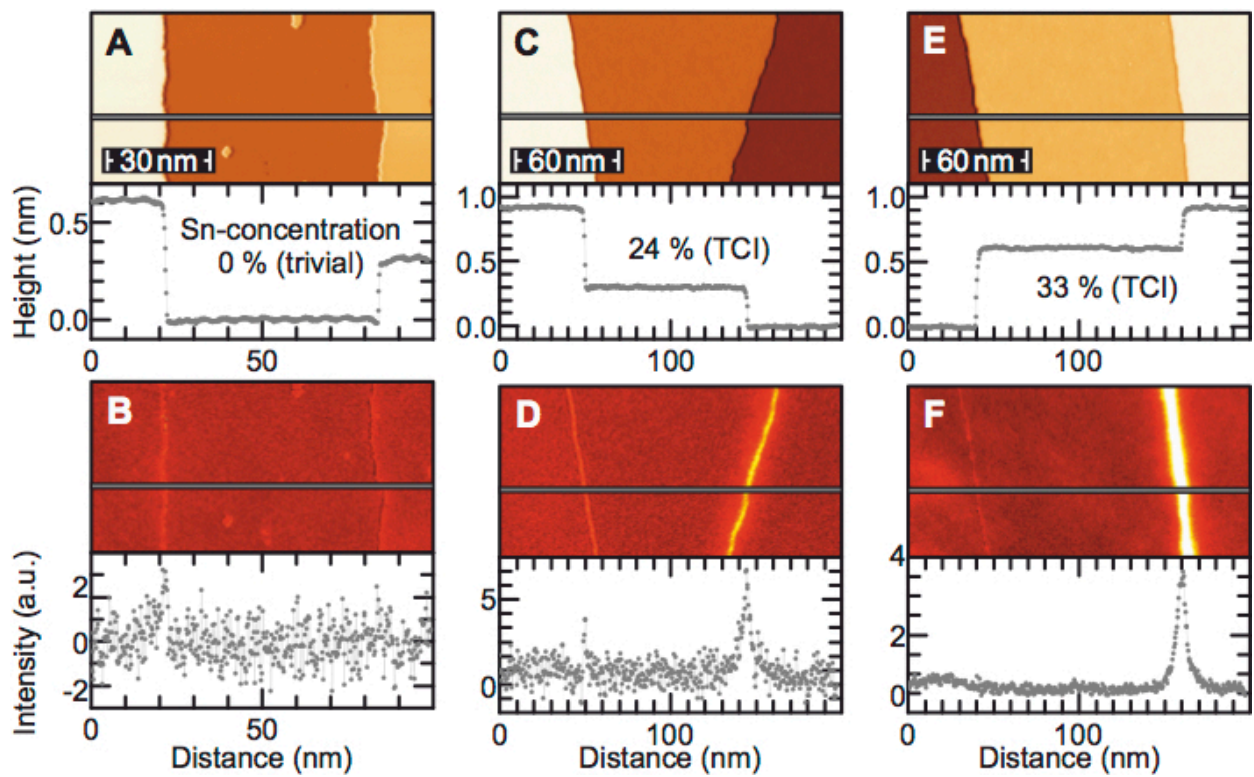

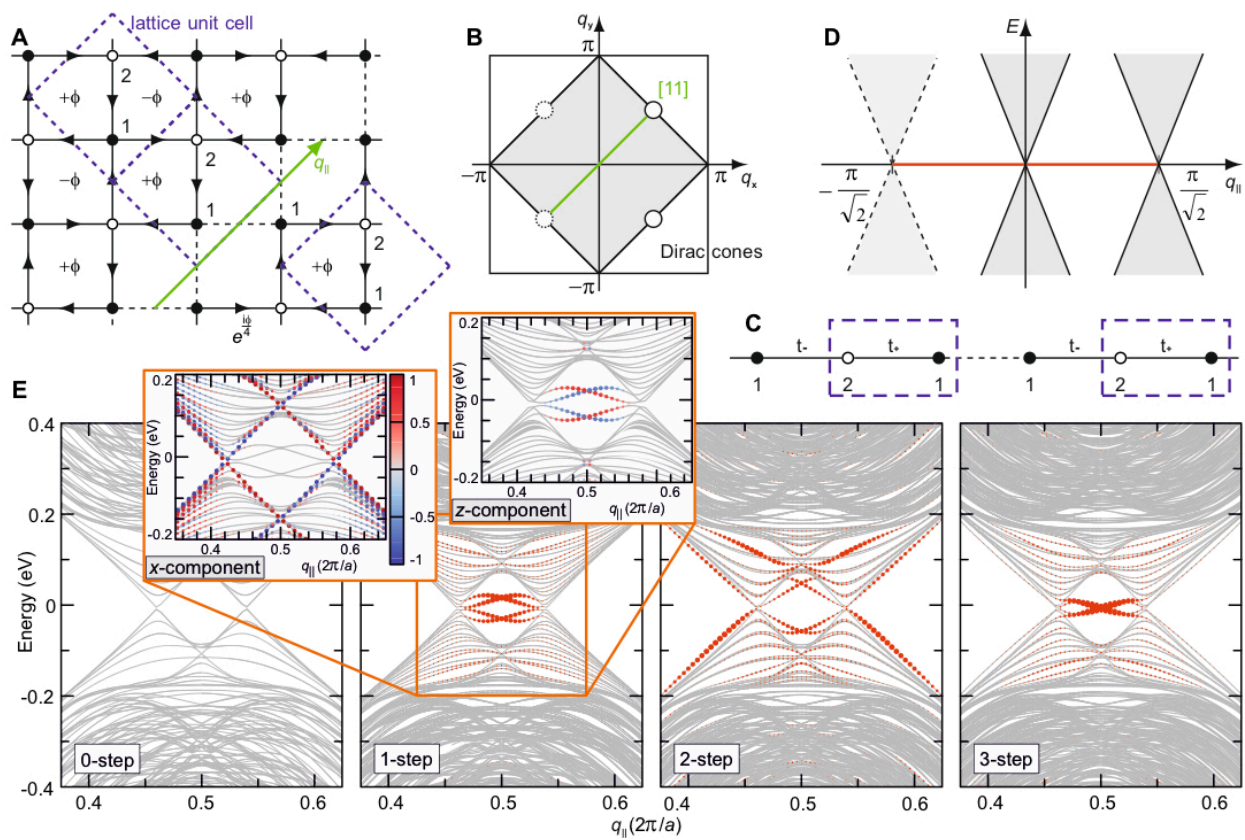

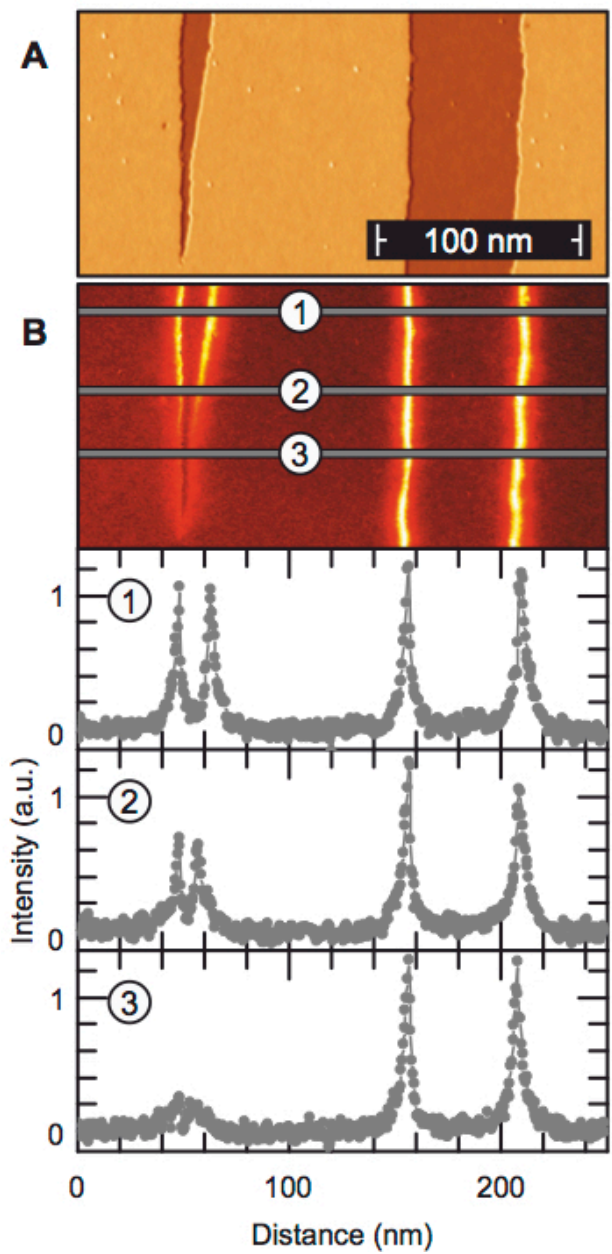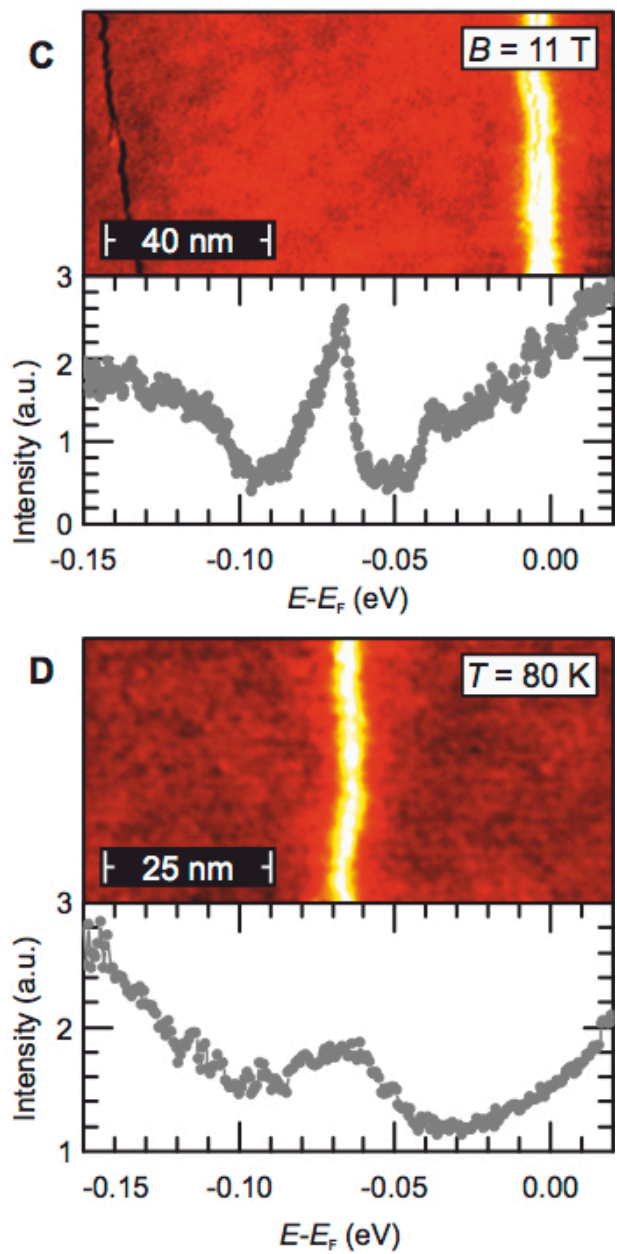